\documentstyle[prd,aps,floats]{revtex}
\begin{document}
\draft

\input epsf \renewcommand{\topfraction}{0.8} 
\newcommand{\beq}{\begin{equation}}
\newcommand{\eeq}{\end{equation}}
\newcommand{\pbar}{\not{\!\partial}}
\newcommand{\dbar}{\not{\!{\!D}}}
\def\lsim{\:\raisebox{-0.75ex}{$\stackrel{\textstyle<}{\sim}$}\:}
\def\gsim{\:\raisebox{-0.75ex}{$\stackrel{\textstyle>}{\sim}$}\:}
\twocolumn[\hsize\textwidth\columnwidth\hsize\csname 
@twocolumnfalse\endcsname

\title{Hubble-induced radiative corrections and Affleck-Dine baryogenesis}   
\author{$^{\dagger}$Rouzbeh Allahverdi,~$^{\dagger}$Manuel
Drees~and~$^{\sharp}$Anupam Mazumdar}     
\address{$^{\dagger}$ Physik Department, TU M\"unchen, James Frank
Strasse, D-85748, Garching, 
Germany. \\
$^{\sharp}$ The Abdus Salam International Center for Theoretical Physics,~I-34100~Trieste, Italy.}
\date{\today} 
\maketitle
\begin{abstract}
We examine the viability of the Affleck-Dine mechanism for
baryogenesis under radiatively induced running of soft breaking $({\rm
mass})^2$ of the flat directions stemming from non-zero energy density
of the inflaton during inflation. A major difference from analogous
phenomenological studies is that the horizon radius provides a natural
infrared cut-off to the quantum corrections in this case. We identify
different scenarios which may arise and point out that the $H_{u}L$
flat direction remains the most promising flat direction, since it is
largely independent of uncertainties about high scale physics and
details of the inflationary model.
\end{abstract}

\pacs{PACS numbers: 12.60.Jv;11.30.Fs;98.80.Cq \hspace*{1.3cm} TUM-HEP-439/01/} 

\vskip2pc]

\section{Introduction}
The Affleck-Dine (AD) mechanism \cite{ad} provides an elegant model
for generating the observed Baryon Asymmetry of the Universe (BAU) in
the framework of supersymmetry; alternative scenarios include GUT
baryogenesis \cite{gutb}, electroweak baryogenesis \cite{krs} and
leptogenesis \cite{fy}. In this scenario some squarks and/or sleptons
acquire a large Vacuum Expectation Value (VEV) along a flat direction
of the scalar potential of the Minimal Supersymmetric Standard Model
(MSSM) during an inflationary epoch (for reviews, see \cite{infl}). A
baryon (or lepton) number violating operator induced by new physics at
a high scale and a large (spontaneously) C and CP violating phase,
provided by the initial VEV along the flat direction, together with
the out of equilibrium condition after inflation, satisfy all three
requirements for the generation of a baryon asymmetry
\cite{sakharov}. The ``AD field'' describing the flat direction starts
oscillating once its mass exceeds the Hubble expansion rate $H$. At
the same time some baryon and/or lepton number violating operator
produces a torque which leads to a spiral motion of the real and
imaginary parts of the VEV in the complex plane. This results in a
baryon (lepton) asymmetry once the comoving number density of the AD
particles is frozen at sufficiently late times \cite{ad}.

In the early Universe the non-zero energy density of the inflaton
field is the dominant source of supersymmetry breaking. This has an
important consequence in models of local supersymmetry where scalar
fields generally acquire a soft supersymmetry breaking $({\rm mass})^2$
component (called Hubble-induced from now on) proportional to $H^2$
\cite{sbi,drt1,drt2}. The effect of such a mass term crucially depends
on the size and sign of the constant of proportionality. A positive
$({\rm mass})^2 \ll H^2$ will not change the analysis of the original
scenario \cite{ad}. On the other hand, for a $({\rm mass})^2 \geq H^2$
the flat direction settles at the origin during inflation and hence
cannot be used to generate BAU. It has been shown that the AD
mechanism leads to interesting amounts of BAU only for a 
$({\rm mass})^2 < {9/ 16} H^2$ \cite{mcdonald}. Perhaps the 
most interesting case occurs for a $({\rm mass})^2 \sim -H^2$,
since it naturally leads to a non-zero VEV of the flat direction 
before the onset of its oscillations. This can be realized at the
tree-level in simple extensions of minimal supergravity 
models \cite{drt1,drt2}, and from one-loop corrections to 
the K\"ahler potential in no-scale supergravity models \cite{gmo}.

A detailed examination of the scenario with $({\rm mass}) ^2 \sim
-H^2$, including a systematic treatment of nonrenormalizable
superpotential terms which lift the flat direction, has been performed
in Ref. \cite{drt2}. Let us denote the AD field describing a generic
direction in the scalar potential of the MSSM which is $D$- and
$F$-flat at the renormalizable superpotential level\footnote{$D$-flat
directions of the MSSM are classified by gauge invariant monomials of
the scalar fields of the theory, $\prod_{i=1}^N \varphi_i$; the AD
field is then defined as the linear combination $\phi \equiv \left(
\sum_{i=1}^N \varphi_i \right) / \sqrt{N}$. For a detailed discussion
of this, as well as the lowest-dimensional operator in the
superpotential which can lift a specific flat direction, see
Ref. \cite{gkm}.} by $\phi$. This flat direction is lifted by a term
in the superpotential of the form
\beq \label{nonren}
W \supseteq {{\lambda}_{n} {\Phi}^n \over n {M}^{n-3}}~,
\eeq
where $\Phi$ is the superfield comprising $\phi$ and its fermionic
partner, $M$ is the scale of new physics which induces the above term,
and ${\lambda}_{n}$ is an ${\cal O}(1)$ number. Supersymmetry breaking
by the inflaton energy density and by the hidden sector result in the
terms
\begin{eqnarray} \label{scalpot}
&-&C_I H^2 {| \phi |}^2 + \left(a {\lambda}_n H {{\phi}^n \over n
M^{n-3}} + {\rm h.c.}\right) + m^{2}_{\phi,0} {| \phi |}^2 \, \nonumber \\ 
&&+ \left(A_{\phi,0} {\lambda}_n {{\phi}^n \over n M^{n-3}} + {\rm h.c.}
\right) \,
\end{eqnarray}
in the scalar potential. The first and the third terms are the
Hubble-induced and low-energy soft mass terms respectively, while the
second and the fourth terms are the Hubble-induced and low-energy $A$
terms respectively. The Hubble-induced soft terms typically dominate
the low-energy ones for $H > m_0$, where 
$m_0 \sim m_{\phi,0} \sim {\cal O}(\rm TeV)$. If $C_I > 0$, the absolute 
value of the AD field during inflation settles at the minimum
given by
\beq \label{veveq}
|\phi| \simeq \left({C_I \over (n-1) {\lambda}_n} H_I M^{n-3}\right)^{1/ n-2} 
\,,
\eeq
with $H_I$ being the Hubble constant during the inflationary
epoch.\footnote{We have ignored the term $\propto a$ in
eq.(\ref{veveq}). If $C_I > 0$, the $a-$term will not change the vev
qualitatively. On the other hand, even for $C_I < 0$ the potential
(\ref{scalpot}) will have a minimum at a nonvanishing vev if $|a|^2 > 4
(n-1) C_I$. However, the origin will also be a minimum in this
case. The viability of the AD mechanism then depends on which minimum
the AD field will ``choose'' during inflation. Because of this
complication we do not pursue the case with $C_I < 0$ and large $|a|$
any further.} If $|a|$ is ${\cal O}(1)$, the phase $\theta$ of
$\langle \phi \rangle$ is related to the phase of $a$ through $n
\theta + {\theta}_a = \pi$; otherwise $\theta$ will take some random
value, which will generally be of ${\cal O}(1)$. After inflation,
$\langle \phi \rangle$ initially continues to track the instantaneous
local minimum of the scalar potential, which can be derived by
replacing $H_I$ with $H(t)$ in Eq.(\ref{veveq}). Once $H \simeq m_0$,
the low-energy soft terms take over. Then the $({\rm mass})^2$ of
$\phi$ becomes positive and $\langle \phi \rangle$ moves in a
non-adiabatic way, since the phase of $\langle \phi \rangle$ during
inflation differs from the phase of $A$. As a result $\langle \phi
\rangle$ starts a spiral motion in the complex plane, which leads to
generation of a net baryon and/or lepton asymmetry
\cite{drt2}. Recently it has been noticed that various thermal effects
from reheating can be substantial which might trigger the motion of
the flat direction at an earlier time and change the yielded BAU
\cite{ace,and}. Detailed studies of AD leptogenesis have been done which
take these thermal effects into account \cite{yanagida}.

\setcounter{footnote}{0} 

All fields which have gauge or Yukawa couplings to the AD field
contribute to the logarithmic running of its $({\rm mass})^2$.
Therefore, one should study the evolution of the flat direction $({\rm
mass})^2$ from some higher scale such as $M_{\rm GUT} $\footnote{We
conservatively choose $M_{\rm GUT} \simeq 2 \cdot 10^{16}$ GeV as the
scale where SUSY breaking is transmitted to the visible sector, in
order to avoid uncertainties about physics between $M_{\rm GUT}$ and
$M_{\rm Planck}$. We further notice that in M-theory scenarios the GUT
scale also represents the string scale\cite{hw}.} down to low scales
in order to determine the location of the true minimum of the
potential and, ultimately, examine the viability of a given flat
direction for the AD mechanism. The running of low-energy soft
breaking masses has been studied in great detail in the context of
MSSM phenomenology\cite{dm}, in particular in connection with
radiative electroweak symmetry breaking \cite{ewsb}. In this note we
perform similar studies in a cosmological set-up for the AD mechanism.


\section{Scale dependence of the flat direction}

We start with a brief review of the running of the soft breaking
$({\rm mass})^2$ of the MSSM scalars. The one-loop beta functions for
the $({\rm mass})^2$ of the Higgs doublet $H_u$ which couples to the
top quark, the right-handed stop $\tilde{u}_3$, the left-handed
doublet of third generation squarks $\tilde{Q}_3$ and the
$A-$parameter $A_t$ associated with the top Yukawa interaction are
\cite{nilles}
\begin{eqnarray} \label{betafct}
{d \over dq} m^2_{H_u} &=& {3h^2_t \over 8 \pi^2}
\left(m^2_{H_u} + m^2_{\tilde{Q}_3} + m^2_{\tilde{u}_3} +
|A_t|^2 \right)  \, \nonumber \\
&-& {1 \over 2 \pi^2} \left({1 \over 4} g^2_1 |m_1|^2 + {3 \over 4}
g^2_2 |m_2|^2 \right) \,, \nonumber \\
{d \over dq}m^2_{\tilde{u}_3} &=& {2h^2_t \over 8 \pi^2}
\left(m^2_{H_u} + m^2_{\tilde{Q}_3} + m^2_{\tilde{u}_3} + |A_t|^2
\right) \ \nonumber \\
&-& {1 \over 2 \pi^2}\left({4 \over 9} g^2_1 |m_1|^2 + {4 \over 3}
g^2_3 |m_3|^2 \right) \,, \nonumber \\
{d \over dq} m^2_{\tilde{Q}_3} &=& {h^2_t \over 8 \pi^2}
\left(m^2_{H_u} + m^2_{\tilde{Q}_3} + m^2_{\tilde{u}_3} + |A_t|^2 \right)\, 
\nonumber \\
&-& {1 \over 2 \pi^2}\left({1 \over 36} g^2_1 |m_1|^2 + {3 \over 4}
g^2_2 |m_2|^2 + {4 \over 3} g^2_3 |m_3|^2 \right) \, ,
\nonumber \\
{d \over dq} A_t &=& {3 h^2_t \over 8 \pi^2} A_t - {1 \over 2 \pi^2}
\left( {13 \over 36} g^2_1 m_1 + {3 \over 4} g^2_2 m_2 + {4 \over 3}
g^2_3 m_3 \right) \, .
\end{eqnarray}
Here $q$ denotes the logarithm of the scale; this could be an external
energy or momentum scale, but in the case at hand the relevant scale
is set by the VEV(s) of the fields themselves. $h_t$ is the
top Yukawa coupling, while $g_1;g_2;g_3$ and
$m_1;m_2;m_3$ are gauge couplings and soft breaking gaugino masses of
the $U(1)_Y;SU(2);SU(3)$ subgroups respectively. If $h_t$ is the only
large Yukawa coupling (i.e. as long as $\tan\beta$ is not very large),
the beta functions for the $({\rm mass})^2$ of squarks of the first and
second generations and the sleptons only receive significant
contributions from gauge/gaugino loops. A review of these effects can
be found in Ref.\cite{dm}. Here we only mention the main results for
universal boundary conditions, where at $M_{\rm GUT}$ the $({\rm mass})^2$ 
of all scalars is $m^2_0$ and the gauginos have the common
soft breaking mass $m_{1/2}$. For a low value of $\tan\beta =1.65$
\footnote{This value corresponds to the case of maximal top
Yukawa coupling, so called fixed point scenario \cite{fps,fps1}, since
this maximal coupling at the weak scale is approached from a wide
range of choices for $h_t$ at the GUT scale. Such a low value of
$\tan\beta$ is excluded by Higgs searches at LEP \cite{lephiggs},
unless one allows stop masses well above 1 TeV. We nevertheless
include this scenario in our discussion since it represents an extreme
case.},
\beq
m^2_{H_u} \simeq  - \frac{1}{2} m^{2}_{0} - 2 m^{2}_{1/2}          
\eeq
at the weak scale, while $m^2_{\tilde u_3}$ and $m^2_{\tilde Q_3}$
remain positive. The soft breaking $({\rm mass}) ^2$ of the first and
second generations of squarks is $\simeq m^{2}_{0} + (5-7)
m^{2}_{1/2}$, while for the right-handed and left-handed sleptons one
gets $\simeq m^{2}_{0} + 0.1 m^{2}_{1/2}$ and $\simeq m^{2}_{0} + 0.5
m^{2}_{1/2}$, respectively. The important point is that the sum
$m^{2}_{H_u} + m^2_L$, which describes the mass in the $H_u L$ flat
direction, is driven to negative values at the weak scale only for
$m_{1/2} \gsim m_0$. This is intuitively understandable since
Eqs.(\ref{betafct}) have a fixed point solution \cite{fps1}
$m^{2}_{H_u} + m^{2}_{\tilde{u}_3} + m^{2}_{\tilde{Q}_3} = A_t = 0$
when $m_{1/2} =0$.

Similarly one could follow the evolution of the soft breaking terms
when the Hubble-induced supersymmetry breaking is dominant, i.e. for
$H > \cal{O}(\rm TeV)$. However, some differences arise in this
case. For the low-energy supersymmetry breaking case, constraints from
the weak scale (e.g., realization of electroweak symmetry breaking,
and experimental limits on the sparticle masses) give information
about $m^2_0$ and $m_{1/2}$. Together with fine tuning arguments,
these constraints imply that $m^2_0 > 0$ and $m_0;m_{1/2}$ are
$\cal{O}(\rm TeV)$. This is different from the Hubble-induced
supersymmetry breaking case, where $m^2_0$ and $m_{1/2}$ are
determined by the scale of inflation (and the form of the K\"ahler
potential). At low scales the Hubble-induced terms are completely
negligible, because at temperature $T \sim M_W$,  $H \sim {\cal
O}(1)$ eV, and at present the Hubble parameter is $H_0 \sim {\cal
O}(10^{-33})$ eV.

There exists an even more fundamental difference between the two
cases. In Minkowski spacetime the contribution of a given loop to a
beta function freezes at a scale of the order of the mass of the
particle in the loop. In an expanding Universe the horizon radius
$\propto H^{-1}$ defines an additional natural infrared cut-off for
the theory. The reason is that the particle description ceases to be
physically meaningful once the Compton wavelength of a particle
exceeds the horizon radius. The masses of particles which are coupled
to the AD field consist of two parts: a supersymmetry preserving part
proportional to the VEV $\langle \phi \rangle$, and the Hubble-induced
supersymmetry breaking part. The contribution of a given loop to a
beta function should thus be frozen at a scale which is the larger of
$|\langle \phi \rangle|$ and $H$ (recall that $h_t$ and gauge
couplings are close to one). In particular, if the squared mass of the
AD field is positive at very large scales but turns negative at some
intermediate scale $Q_c$, the origin of the AD potential will cease to
be a minimum provided the Hubble parameter is less than $Q_c$. On the
other hand, if $m_\phi^2 < 0$ at the GUT scale, its running should
already be terminated at the scale $|\langle \phi \rangle|$ determined
by Eq.(\ref{veveq}).\footnote{Here we note that the Hubble cut-off
usually plays no role in loop corrections to the inflaton
potential. In most inflationary models the masses of the fields which
may run in the loop are larger than the Hubble expansion during
inflation due to the presence of a finite coupling to the
inflaton. This will happen if the inflaton (time varying) VEV is large
and the couplings are not very small. In those cases, which are
somewhat similar to our case with $C_I > 0$, one could right away
trust the usual loop calculation evaluated in a flat space time
background \cite{dss}.} In the following two subsections we therefore
discuss the cases of positive and negative GUT-scale $({\rm mass})^2$
for the AD field separately.


\subsection{The case with $C_I \approx -1$}

\begin{table} \label{table1}
\caption{The scale $Q_c$ (in GeV) where the squared mass of the AD
field describing the $H_u L$ flat direction changes sign, for $C_I =
-1$ and several values for the ratios $A_t/H$ and $m_{1/2}/H$ as well
as the top Yukawa coupling $h_t$, all taken at scale $M_{\rm GUT} = 2
\cdot 10^{16}$ GeV.}

\vspace*{3mm}
\begin{tabular}{|c|c||c|c|}
$A_t/H$ & $m_{1/2}/H$ & $Q_c(h_t = 2)$ & $Q_c(h_t = 0.5)$ \\
\hline
$ +1/3$ ($-1/3$)& $1/3$ & $\times $ & $\times $ \\
$+1/3$ ($-1/3$)& $1$ & $10^6-10^7$ & $10^3$ \\
$+1/3$ ($-1/3$)& $3$ & $10^{11}$ & $10^6-10^7$ \\
\hline
$+1$ ($-1$)& $1/3$ & $\times$  & $\times$  \\
$+1$ ($-1$)& $1$ & $10^6-10^7$ & $10^5$ ~ ($\times$) \\
$+1$ ($-1$)& $3$ &  $10^{11}$ &  $10^8$ ~ ($10^6$) \\
\hline
$+3$ ($-3$)& $1/3$ & $\times$  & $10^7$ ~  \\
$+3$ ($-3$)& $1$ & $10^{14}$ ~ ($10^7$) & $10^9$ ~ ($10^3$) \\
$+3$ ($-3$)& $3$ & $10^{15}$ ~ ($10^{11}$) & $10^{10}$ ~ ($10^6$)
\end{tabular}
\end{table}
In this case all scalar fields roll towards the origin very rapidly
and settle there during inflation if radiative corrections to their
$({\rm mass})^2$ are negligible. A typical AD field $\phi$ is a linear
combination $\phi = \sum_{i=1}^N a_i \varphi_i$ of the MSSM scalars
$\varphi_i$, implying that $m^{2}_{\phi} = \sum_{i=1}^N |a_i|^2
m^{2}_{\varphi}$. As mentioned before, the running of $m^{2}_{\phi}$
crucially depends on $m_{1/2}$. A Hubble-induced gaugino mass can be
produced from a (non-minimal) dependence of the gauge superfield
kinetic terms on the inflaton field.  Generally the gauge superfield
kinetic terms must depend on the field(s) of the hidden or secluded
sector in order to obtain gaugino masses of roughly the same order as
(or larger than) scalar masses, as required by phenomenology. Having
$m_{1/2} \sim H$ thus appears to be quite natural unless an
$R$-symmetry forbids terms which are linear in the inflaton superfield
\cite{drt2}. The same also holds for the Hubble-induced $A$ terms. The
$\mu$ term is a bit different. Since it doesn't break supersymmetry,
there is a priori no reason to assume that $\mu$ of order $H$ will
be created. However, it seems more appealing to evoke some mechanism
\cite{muprob1,muprob2} that naturally produces $\mu$ of order of the
soft breaking masses in Minkowski space. A $\mu-$term of order $H$
can probably be realized in the models of Ref.~\cite{muprob1}, but
seems unlikely to emerge in those of Ref.~\cite{muprob2}. We will
therefore treat $\mu$ as a free parameter. We will see below that
small values of $\mu$ are favored.

We considered sample cases with $m_{1/2} = H;3H;H/3$, $A_t(M_{\rm
GUT}) =\pm H;\pm 3H;\pm H/3$\footnote{The RGE (\ref{betafct}) for
$A_t$ shows that the relative sign between $A_t$ and $m_{1/2}$ matters
since it affects the running of $|A_t|$ and, subsequently, scalar soft
masses. Without loss of generality we take the common gaugino mass
$m_{1/2}$ to be positive.}, $h_t(M_{\rm GUT}) = 2, \, 0.5$ and
$g_1(M_{\rm GUT})=g_2(M_{\rm GUT})=g_3(M_{\rm GUT})=0.71$. We then
followed the running of scalar soft masses from $M_{\rm GUT}$ down to
$10^3$ GeV, where the low-energy supersymmetry breaking becomes
dominant.

The main observation is that only the $H_{u}L$ flat direction can
acquire a negative $({\rm mass})^2$ at low scales. In this case
$m^2_{\phi} = {(m^2_{H_u} + m^2_{L} + \mu^2)/ 2}$, where the last term
is the contribution from the Hubble-induced $\mu$ term. The results
for this case are summarized in TABLE I, for $\mu (M_{\rm GUT}) \lsim
H/4$, so that the contribution $\propto \mu^2$ to $m^2_\phi$ is
negligible. In general $m^2_\phi$ changes sign at a higher scale for
$h_t(M_{\rm GUT}) = 2$. This is expected since a larger Yukawa
coupling naturally maximizes the running of $m^2_{H_u}$. Furthermore,
the difference between $A_t/m_{1/2} < 0$ and $A_t/m_{1/2} > 0$ becomes
more apparent as $|A_t/m_{1/2}|$ increases and $h_t$ decreases. The
quasi fixed-point value of $A_t/m_{1/2}$ is positive
\cite{fps1}. Positive input values of $A_t$ will thus lead to positive
$A_t$ at all scales, but a negative $A_t(M_{\rm GUT})$ implies that
$A_t \simeq 0$ for some range of scales, which diminishes its effect
in the RGE, see Eq.~(\ref{betafct}). The sign of $A_t(M_{\rm GUT})$ is
more important for smaller $h_t$, since then $A_t/m_{1/2}$ will evolve
less rapidly.

We also notice that the squared mass of the $H_u L$ flat direction
does not change sign when $m_{1/2} = H/3$, except for $A_t = \pm 3H$
and $h_t = 0.5$\footnote{For this choice of parameters, $A_t$
initially runs very slowly. It will therefore remain large for some
time and helps $m^{2}_{H_u}$ to decrease quickly towards lower
scales.}. This can be explained by the fact that for small $m_{1/2}$
and small or moderate $|A_t|$ we are generally close to the fixed
point solution
\beq \label{fixeq}
m^2_{H_u} \simeq -{1 \over 2} H^2; ~m^2_{\tilde{u}_3} \simeq
0; ~m^2_{\tilde{Q}_3} \simeq {1 \over 2} H^2.
\eeq
Nevertheless, even for $m_{1/2} \ll H$ the squared mass of the
$H_{u}L$ flat direction as well as $m^2_{\tilde{u}_3}$ are $< 0.2
H^2$ well above 1 TeV, exactly due to the fixed point solution
behavior. This implies that the $H_{u}L$ flat direction can still be
viable \cite{mcdonald}. Flat directions built out of $\tilde u_3$
will be marginal at best, since the reduction of $m^2_{\tilde u_3}$
will be diluted by other contributions to $m^2_\phi$ that are not
reduced by RG running; e.g. for the $U_3 D_1 D_2$ flat direction we
find $m^2_\phi > 2 H^2 / 3$ at all scales.

The AD mechanism should always work if $Q_c > H_I$, since then the
global minimum of the potential during inflation is located at
$|\langle \phi \rangle | \not= 0$. Note that in this case the vev
$|\langle \phi \rangle|$ is usually determined by $Q_c$ rather than by
Eq.~(\ref{veveq}). For scales close to $Q_c$ the mass term in the
scalar potential Eq.~(\ref{scalpot}) can be written as $\beta_\phi
H^2 |\phi|^2 \log({|\phi|}/{Q_c})$, where the coefficient
$\beta_\phi$ can be obtained from the RGE. If $\beta_\phi > 0$, which
is true for the $H_u L$ flat direction for $C_I < 0$, this term will
reach a minimum at $ \log({|\phi|}/{Q_c}) = -1$. If $Q_c < (H_I
M^{n-3}_{\rm GUT})^{1/n-2}$ the non-renormalizable contributions to
the scalar potential are negligible for $|\phi| \sim Q_c$, so that the
minimum of the quadratic term essentially coincides with the minimum
of the complete potential given by Eq.~(\ref{scalpot}). In models of 
high scale inflation (e.g. chaotic inflation models), the Hubble 
constant during inflation $H_I$ can be as large as $10^{13}$ GeV. 
This implies that $m^2_\phi$ for the $H_u L$ flat direction can 
only become negative during inflation if $m^2_{1/2} \gg H^2$, which 
includes the ``no-scale'' scenario studied in Ref.~\cite{gmo}. The region of
parameter space safely allowing AD leptogenesis is much larger in
models of intermediate and low scale inflation (e.g. some new
inflation models) where $H_I$ is substantially smaller. In such models
one can easily have $H_I < Q_c$ at least for the $H_u L$ flat
direction, unless $m^2_{1/2} \ll H^2$ or $\mu^2 \gsim m^2_{1/2}$.

If $Q_c < H_I$, $\phi$ settles at the origin during inflation and its
post-inflationary dynamics will depend on the process of
thermalization. If the inflaton decay products thermalize very slowly,
$m^{2}_{\phi}$ is only subjected to zero-temperature radiative
corrections and $\langle \phi \rangle$ can move away from the origin
once $H \lsim Q_c$; a necessary condition for this scenario is that
inflatons do not directly decay to fields that are charged under
$SU(3) \times SU(2) \times U(1)_Y$. If $Q_c \gg 1$ TeV, $\phi$ will
readily settle at the new minimum and AD leptogenesis can
work. However, the situation will be completely different if inflatons
directly decay to some matter fields. In such a case the plasma of
inflaton decay products has a temperature $T \sim ({\Gamma}_{d} H
M^{2}_{\rm Planck})^{1/4}$ \cite{kt} (${\Gamma}_{d}$ is the inflaton
decay rate). Thus fields which contribute to the running of $m^2_\phi$
are in thermal equilibrium (recall that the AD field is stuck at $\phi
= 0$) and their back reaction results in thermal corrections of order
$+T^2$ to $m^2_\phi$. For generic models of inflation $T > H$,
implying that thermal effects exceed radiative corrections. Therefore
$\langle \phi \rangle$ remains at the origin at all times and AD
leptogenesis will not work.


\subsection{The case with $C_I \approx +1$}

In this case all flat directions are viable if the running of
$m^{2}_{\phi}$ is negligible. However, radiative corrections may
change the sign (in this case to positive) at small vev(s) possibly
resulting in the entrapment of $\phi$ at the origin. We quantitatively
studied the same sample cases as above. 

The main results can be summarized as follows. The squared mass of the
AD field for the $H_{u}L$ flat direction is always negative at small
scales, unless $\mu^2 \gsim H^2/2$. However, for $m_{1/2} = 3H$,
$m_\phi^2$ changes sign twice; it is positive for scales $Q$ between
roughly $10^{14}$ and $10^6$ GeV, the precise values depending on
$h_t$ and $A_t$. Slepton masses only receive positive contributions from
electroweak gauge/gaugino loops. As a result, the squared mass of the
AD field describing the $LLE$ flat direction remains negative down to
1 TeV, unless $m_{1/2} > 2 H$; for $m_{1/2} \gsim 3 H$, $Q_c \gsim
10^9$ GeV even for this flat direction. The squared masses of all
squarks (except $\tilde{u}_3$) change sign at $Q_c > 1$ TeV unless
$m_{1/2} \lsim H/3$; we find $Q_c \simeq 10^{10} \ (10^{15})$ GeV for
$m_{1/2} / H = 1 \ (3)$. This is due to the large positive
contribution $\propto m_3^2$ to the squared squark masses at scales
below $M_{\rm GUT}$. The corresponding values for the $U_3 D_i D_j$
and $L Q D$ flat directions are usually somewhat smaller, due to the
Yukawa terms in the $\beta-$function and the slower running of slepton
masses, respectively; however, the listed values of $Q_c$ are still a
fair approximation for these cases.

According to Eq.(\ref{veveq}), the scale $ |\langle \phi \rangle| \gg H$, 
above which the positive contribution to the scalar potential from
the nonrenormalizable superpotential term in Eq.~(\ref{nonren}) dominates
$-H^2$, now appears. If $Q_c > (H_I M^{n-3}_{\rm GUT})^{1/n-2}$,
$m^2_\phi$ is positive for all vev(s) and hence the flat direction
will settle at the origin during inflation and remain there from then
on. In such a case the flat direction is not viable for the AD
mechanism. This can easily happen for flat directions involving
squarks in models of low scale inflation, but is not likely for high
scale inflationary models unless $m_{1/2} \gsim 3H$. For 
$H_I < Q_c <(H_I M^{n-3}_{\rm GUT})^{1/n-2}$, feasible for some 
flat directions in both intermediate/high scale and low scale models, 
the potential during inflation has two minima, at $\langle \phi \rangle = 0$ 
and at $|\langle \phi \rangle | \sim (H_I M^{n-3}_{\rm GUT})^{1/n-2}$. 
Depending on the initial conditions, $\phi$ can roll
down towards either of them and settle there but only the latter one
will be useful for the AD mechanism. If $Q_c < H_I$, the AD field
direction will settle at the value determined by Eq.~(\ref{veveq}) 
(the only minimum during inflation) and remain there afterwards. 
The appearance of another minimum at the origin after inflation, 
which is possible once $H < Q_c$, does not change the situation 
since these minima are separated by a barrier. Therefore in this 
case radiative corrections will not change the picture qualitatively; 
however, they will still modify the quantitative analysis, 
since $C_I$ in Eq.(\ref{veveq}) will become scale-dependent.

In summary, for models of high/intermediate scale inflation the AD
mechanism will not be disrupted unless $m_{1/2} \gsim 3H$. On the
other hand, the $H_{u} L$ flat direction is the most promising one for
low scale inflationary models, regardless of the value of $m_{1/2}$,
provided only that the Hubble-induced $|\mu|$ is not too
large. Similarly, if $m_{1/2} \gsim 3H$, AD leptogenesis along the
$H_u L$ flat direction is the only viable option, but requires a
relatively low scale $H_I$. However, $Q_c \ll {(H_I
M^{n-3}_{\rm GUT})}^{1/2}$ and $|\langle \phi \rangle | \sim Q_c$ at
the minimum of potential in this case. Thermal effects may therefore
trigger an early oscillation of the flat direction, if inflaton
directly decays to matter fields \cite{ace,and}.   


\section{Conclusion}

In this note we examined the AD mechanism for baryogenesis including
radiative corrections to the Hubble-induced soft breaking 
$({\rm mass})^2$ of the MSSM scalars. An important point is that in an
expanding Universe the horizon radius provides a natural infrared
cut-off to such corrections. Radiative corrections lead to interesting
consequences whenever the Hubble-induced soft breaking parameters
satisfy $m_{1/2} \gsim m_0$, or $|A_0| \gsim m_0$ with $m_0^2 > 0$;
here $m_0, \, m_{1/2}$ and $A_0$ are the common soft breaking scalar
and gaugino masses and common trilinear soft breaking parameter,
respectively, all taken at the input scale $M_{\rm GUT}$. We found
that the $H_u L$ flat direction remains viable for a large region of
parameter space, in particular for both signs of $m_0^2$, as long as
the Hubble-induced $\mu-$parameter satisfies 
$|\mu|^2 \lsim {1}/{4} \max\{m_0^2, m^2_{1/2}\}$. In contrast, 
flat directions involving squarks are only viable for $m_0^2 < 0$ and 
relatively small $m_{1/2}$, the precise upper bound depending on the 
Hubble parameter during inflation $H_I$. Purely sleptonic flat directions are
intermediate between these two extreme cases.

It should be emphasized that the values of $m_0, \ m_{1/2}, \ A_0$ and
$\mu$ used in this analysis have no bearing on present
phenomenology. All these parameters are Hubble-induced, and thus
contribute negligibly to the present-day sparticle spectrum. In
particular, our analysis will go through even if present-day
supersymmetry breaking is not due to gravity mediation, as long as
physics at scales around $M_{\rm GUT}$ can be described by an
effective Supergravity theory in 4 dimensions.  The only MSSM
parameter which is of some importance for our analysis is the ratio of
vevs $\tan\beta$, which determines the Yukawa couplings of the quarks
and leptons. However, we saw in Sec.~IIA that changing the top
coupling at the GUT scale from 0.5 to 2.0 does not lead to large
variations in AD phenomenology. If $\tan\beta \gg 1$, i.e. for large
bottom and $\tau$ Yukawa couplings, the domain of viability for flat
directions involving $\tilde b$ and/or $\tilde \tau$ fields will
increase somewhat, in particular for $m_0^2 > 0$, but again we do not
expect the situation to change qualitatively in this case.

Moreover, we do not find direct consequences for Q-ball \cite{ks,em}
production. Radiative corrections to the Hubble-induced soft mass can
affect the initial conditions at the onset of flat direction
oscillations. On the other hand, Q-ball formation occurs during
oscillations which start when the low-energy supersymmetry breaking
\cite{drt2} or thermal effects \cite{ace} dominate the Hubble-induced
supersymmetry breaking. Once the latter becomes subdominant, the
standard analysis \cite{ks,em} of the flatness of scalar potential and
Q-ball formation will apply. However, there is an indirect connection,
since Q-balls might evaporate if the inflaton decay products
thermalize quickly, unless the reheat temperature is very low. We saw
in Sec.~IIA that models with delayed thermalization might realize AD
baryogenesis with $m_0^2 > 0$ even if $Q_c < H_I$, as long as $Q_c \gg
1$ TeV.

Finally, we reiterate that radiative corrections will not affect the
AD mechanism qualitatively if there exists an $R$-symmetry which
forbids the appearance of terms linear in the inflaton superfield
(which would imply $m_{1/2}, |A_0| \ll m_0$). On the other hand, in
more general scenarios $Q_c$ depends very strongly (essentially
exponentially) on $m_{1/2}$, and is thus very sensitive to details of
physics at high scales. We have seen that AD leptogenesis from the
$H_{u}L$ direction is quite robust and works (almost) independently of
the size of $m_{1/2}$ and the sign of $m^2_0$, as long as the
Hubble-induced $\mu-$term is not too large. Recall that this scenario
also has the distinction of connecting baryogenesis with the neutrino
sector parameters \cite{yanagida}. Our analysis provides an argument
why the $H_{u}L$ direction might be preferred {\em dynamically} over the
plethora of other possible flat directions.


\section*{Acknowledgements}

The authors thank A. Perez-Lor\'enzana for fruitful discussion.
The work of R.A. and M.D. was supported by
``Sonderforschungsbereich 375 f\"ur Astro-Teilchenphysik'' der
Deutschen Forschungsgemeinschaft. A.M. acknowledges the support of
{\bf The Early Universe Network} HPRN-CT-2000-00152.


\end{document}